\documentclass[journal,9pt]{IEEEtran}

\usepackage{amsmath,amsfonts}
\usepackage{algorithmic}
\usepackage{algorithm}
\usepackage{array}
\usepackage{xcolor}
\usepackage{subfigure}
\usepackage{subcaption}
\usepackage{textcomp}
\usepackage{stfloats}
\usepackage{url}
\usepackage{verbatim}
\usepackage{graphicx}
\usepackage{cite}
\usepackage{hyperref}
\usepackage[normalem]{ulem}
\DeclareMathOperator*{\argminA}{arg\,min}
\usepackage{booktabs}
\usepackage{multirow}

\begin{document}

\title{\huge{Computational Microwave Imaging Relying on Orbital Angular Momentum Transmitarrays for Improved Diversity}} 

\author{\Large{Miguel Angel Balmaseda-Marquez,~\IEEEmembership{Graduate Student Member,~IEEE,} Guillermo Álvarez-Narciandi,~\IEEEmembership{Senior Member,~IEEE,} María García-Fernández,~\IEEEmembership{Senior Member,~IEEE,} Carlos Molero Jiménez,~\IEEEmembership{Senior Member,~IEEE,} William Whittow,~\IEEEmembership{Senior Member,~IEEE,} and Okan Yurduseven,~\IEEEmembership{Senior Member,~IEEE}}
\thanks{This work has been supported in part by MICIU/AEI/10.13039/501100011033/FEDER, UE, under grants PID2024-155167OA-I00, PID2024-160008OA-I00, RYC2023-043020-I, RYC2023-045010-I and FPU21/02219, and in part by Consejería de Universidad, Investigación e Innovación of Junta de Andalucía through grant EMERGIA 23-00235. This work was also supported in part by the Leverhulme Trust under the Research Leadership Award under Grant RL-2019-019, and by U.K. Research and Innovation (UKRI) Postdoctoral Fellowship Guarantee for Marie Skłodowska-Curie Action Postdoctoral Fellowships under Project EP/X022951/1 and Project EP/X022943/1. (\textit{Corresponding author: Miguel Angel Balmaseda-Marquez}.)}.
\thanks{Miguel Angel Balmaseda-Marquez is Dept. of Signal Theory, Telematics and Communications and with CITIC-UGR, University of Granada, Spain (e-mail: migbalmar@ugr.es)}
\thanks{Carlos Molero Jiménez with Dept. of Electronic and Electromagnetism, Faculty of Physics, University of Seville, 41012, Seville, Spain (e-mail: cmolero1@us.es).}
\thanks{Guillermo Álvarez-Narciandi and María García-Fernández are with the Group of Signal Theory and Communications, University of Oviedo, 33003 Gijón, Spain (e-mail: alvareznguillermo@uniovi.es; garciafmaria@uniovi.es).} 
\thanks{William Whittow is with the Wolfson School of Mechanical, Electrical and Manufacturing Engineering, Loughborough University, LE11 3TU Loughborough, U.K (e-mail: w.g.whittow@lboro.ac.uk).} 
\thanks{Okan Yurduseven is with the Centre for Wireless Innovation, School of Electronics, Electrical Engineering and Computer Science, Queen's University Belfast, BT9 5BN, Belfast, Northern Ireland, UK (e-mail: okan.yurduseven@qub.ac.uk).}
\thanks{Manuscript received November 21, 2025; revised XXXX; accepted XXXX.}}

\markboth{IEEE Transactions on Antennas and Propagation,~Vol.~XX, No. XX, 2025}%
{Shell \MakeLowercase{\textit{et al.}}: A Sample Article Using IEEEtran.cls for IEEE Journals}


\maketitle

\begin{abstract}
This work proposes the use of orbital angular momentum (OAM) waves to improve the performance of a computational imaging (CI) system. Specifically, in contrast to a solely frequency-diverse operation, leveraging multiple OAM waves leads to a significant increase in the diversity of the measurement modes of a CI system. 
This significantly reduces the frequency bandwidth required to achieve high-quality image reconstructions. A proof-of-concept prototype working at Ka-band frequencies is used to validate the proposed approach. The prototype consists of two metalized three-dimensional (3D) printed cavities, with fully-dielectric transmitarrays inside that generate OAM waves. Imaging results from various targets reveal that the CI system achieves superior imaging quality when multiple OAM waves are considered, compared to when it solely relies on frequency-diversity. This is specially noticeable in the case of complex distributed targets, which can only be reconstructed with the prototype when multiple OAM waves are used. Furthermore, it is shown that accurate image reconstructions can be obtained employing only one eighth of the operational bandwidth of the frequency-diverse system.

\end{abstract}


\section{Introduction}
\IEEEPARstart{M}{icrowave} radar-based imaging and sensing systems have been extensively studied due to their wide range of applications, including remote sensing \cite{He2023-RS,Zhang2022-RS,Jonard2018-RS,Ma2022-RS,Montzka2016-RS}, biomedical imaging \cite{Qanoune2024-BI, Meaney2012-BI, Mojabi2014-BI ,Bialkowski2016-BI}, and security screening \cite{Zhuravlev2023-SS,Ahmed2021-SS,Salmon2020-SS}. This interest arises from several advantages of electromagnetic (EM) waves at microwave frequencies, such as their non-ionizing nature, their ability to operate under various weather conditions, and their capability to penetrate several optically opaque materials. A comprehensive review of microwave imaging techniques in the literature reveals that many studies rely on mechanical or electronic raster scanning methods. For instance, in synthetic aperture radar (SAR) systems, the movement of the radar platform is commonly employed for scene reconstruction \cite{PSM_dron,Guille2019-SAR}.


Unlike the aforementioned approaches, the use of computational imaging (CI) systems was proposed to bypass the need of rastering the scene under inspection \cite{Hunt_comp_imaging}. CI systems employ a set of spatially incoherent radiation patterns (also noted as \emph{measurement modes}) to illuminate the scene, achieving a physical-layer compression of its information. During the image reconstruction process, this information can be decoded using various signal processing techniques \cite{Lipworth2015, Okan_screening}. CI systems are usually classified according to how the different measurement modes are generated. Frequency-diverse CI systems generate frequency-distinct radiation patterns to probe the scene \cite{TWI_CI,Okan_screening}. Thus, they rely on the use of large operational bandwidths (BWs) to synthesize a sufficient number of measurement modes. Alternatively, other CI systems resort to the dynamic reconfiguration of the transmitter (TX) and the receiver (RX), easing the large-bandwidth requirement. An example of this type of CI system can be implemented using dynamic metasurface antennas (DMAs) \cite{DMA_AWPL,TiejunCui_RIS_CI}. In that case, reconfigurable elements, such as diodes, are tuned to reconfigure the radiation pattern of the antenna. A hybrid approach, in which the frequency-diversity and the dynamic reconfiguration principles are exploited can also be adopted \cite{review_Imani}.

As previously explained, CI systems rely on the use of spatially incoherent measurement modes to probe the scene under inspection. However, in practice, there is always a certain degree of correlation among modes. Thus, a critical aspect of CI systems is the degree of correlation between measurement modes (i.e., how diverse are the measurement modes). A greater diversity enables the system to encode the scene information with fewer modes as the information redundancy decreases (e.g., a reduced bandwidth operation becomes possible). Towards improving the diversity of the measurement modes of CI systems, this work proposes to take advantage of the properties of angular orbital momentum (OAM) waves.


OAM waves, also known as vortex waves, are characterized by wavefronts with helically-distributed phase \cite{Griffiths-OAM, Allen1992-OAM}.
In contrast to waves with spin angular momentum (SAM), mainly associated with circularly-polarized waves, an OAM wave can carry \textit{L} orders that take arbitrary integer values ($l$ = -1, 0, 1...). Different OAM eigenstates are mutually orthogonal to each other in each frequency, which makes it a promising physical dimension to be exploited in optical and microwave systems \cite{Yu2016-OAM}. Vortex waves have demonstrated several advantages in object recognition and remote sensing applications, including the use of digital spiral imaging techniques based on OAM modes \cite{Zhang2018-OAM}. Furthermore, radio vortex imaging has been proposed as an effective approach for acquiring azimuthal information in two-dimensional (2D) radar target systems \cite{Liu2018-OAM,Liu2020-OAM}. 

There are several methods for generating OAM waves \cite{Zhang2020-OAM,Wang2024-OAM}. At higher frequencies, including the optical domain, spiral phase plates are commonly used to produce vortex waves \cite{Hui2015-OAM}. Dielectric plate structures with spatially varying permittivity can also generate specific OAM beams \cite{Chen2016-OAM}. Additionally, circular traveling-wave antennas based on ring resonant cavities and classical devices such as uniform circular arrays have been employed to produce OAM beams at millimeter-wave frequencies \cite{Zheng2015-OAM, Guo2017-OAM}. More recently, dielectric metasurfaces have emerged as an efficient, cost-effective, and straightforward approach for generating OAMs \cite{Balma2024-OAM,Lin2022-OAM}. Advances in three-dimensional (3D) printing have further improved fabrication precision, enabling the production of high-frequency metasurfaces with reduced manufacturing tolerances like those seen in \cite{Moreno2022-UC, Huang2020-SLA}

This work shows that the use of OAM waves can significantly increase the diversity of a frequency diverse CI system. The orthogonal behaviour of the OAM modes enables a narrow band frequency operation. To analyze this, a proof-of-concept CI system, comprising a 3D printed metallized mode-mixing cavity with an integrated transmitarray (TA) generating OAM modes was designed, fabricated and measured. The results reveal that the use of multiple OAM modes produces higher quality image reconstructions. The rest of the paper is organized as follows: Section \ref{sec:CI} shows the CI paradigm, Section \ref{sec:system_description} presents the proposed approach and the proof-of-concept CI system, Section \ref{sec:results} discusses the results and Section \ref{sec:conclusion} draws the conclusion.

\section{Computational Imaging}
\label{sec:CI}
The CI paradigm relies on the use of antennas that are capable of radiating a set of spatially incoherent radiation patterns. As a result, CI systems can probe the scene under inspection from a single acquisition position (or a reduced number of them), overcoming the demanding sampling requirements of traditional SAR systems. This yields a significant reduction of the overall hardware complexity of the imaging system, but there is no longer a one-to-one relationship between the radiating elements and the measured signal, which needs to be \emph{decompressed} taking into consideration the patterns radiated by the TX/RX antennas \cite{review_Imani,vasiliki_freqDomain,amir_kirchoff}.

Taking into account the first Born approximation \cite{Lipworth2015}, the received backscattered radar measurements, $\mathbf{g}$, can be expressed as

\begin{equation}
\label{eq:CI}
\mathbf{g}_{M\times1} = \mathbf{H}_{M\times N} \mathbf{\sigma}_{N\times1} + \mathbf{n}_{M\times1}
\end{equation}
where $\mathbf{H}$ is the sensing matrix of the CI system, which is proportional to the fields radiated by the transmit and receive apertures, $\mathbf{E}_{\textrm{TX}}$ and $\mathbf{E}_{\textrm{RX}}$, respectively, propagated to the imaged scene domain \cite{TWI_CI}. The reflectivity of the scene under inspection, which is discretized into $N$ voxels, is noted by $\mathbf{\sigma}$, and $\mathbf{n}$ is a  noise term. The number of measurement modes of the CI system is represented by $M$, which in a frequency diverse system is equal to the number of acquired frequency samples, $N_\text{f}$. In this work, as the diversity of the radiation patterns generated by a mode-mixing cavity is increased considering several OAM waves, $M=N_\text{f}\cdot N_{\text{OAM}}$, where $N_{\text{OAM}}$ refers to the number of considered TX-RX OAM wave combinations. In particular, during the acquisition process a frequency sweep is performed for each OAM configuration. Finally, an estimate of the reflectivity of the scene can be retrieved using a least squares minimization:

\begin{equation}
\label{eq:LS}
\mathbf{\sigma}_{\text{est}} = \argminA_\sigma ||\mathbf{g}-\mathbf{H}\mathbf{\sigma} ||^2_2
\end{equation}
where $||\cdot||$ is the Euclidean norm.

\section{System Description}
\label{sec:system_description}
The imaging system model presented here consists of a leaky-front cavity with a random distribution of holes, stimulated by OAM-waves with different orders $l$. In this work the maximum OAM-wave order considered is $l=+10$, but the value of $N_{\text{OAM}}$ can be increased using different OAM waves in the TX and the RX. The OAM-waves are generated by internal fully dielectric TAs. To demonstrate the image reconstruction enhancement that can be achieved using different OAM waves, a proof-of-concept prototype, which is manually reconfigured by changing the dielectric TA inside the cavity, was manufactured. For this purpose, a set of static TAs was designed and fabricated to induce a set of mono-order OAM excitations.



\subsection{OAM Transmitarrays}
\label{sec:TAsdesign}

\begin{figure}[]
    \centering
    {\includegraphics[width=0.92\columnwidth]{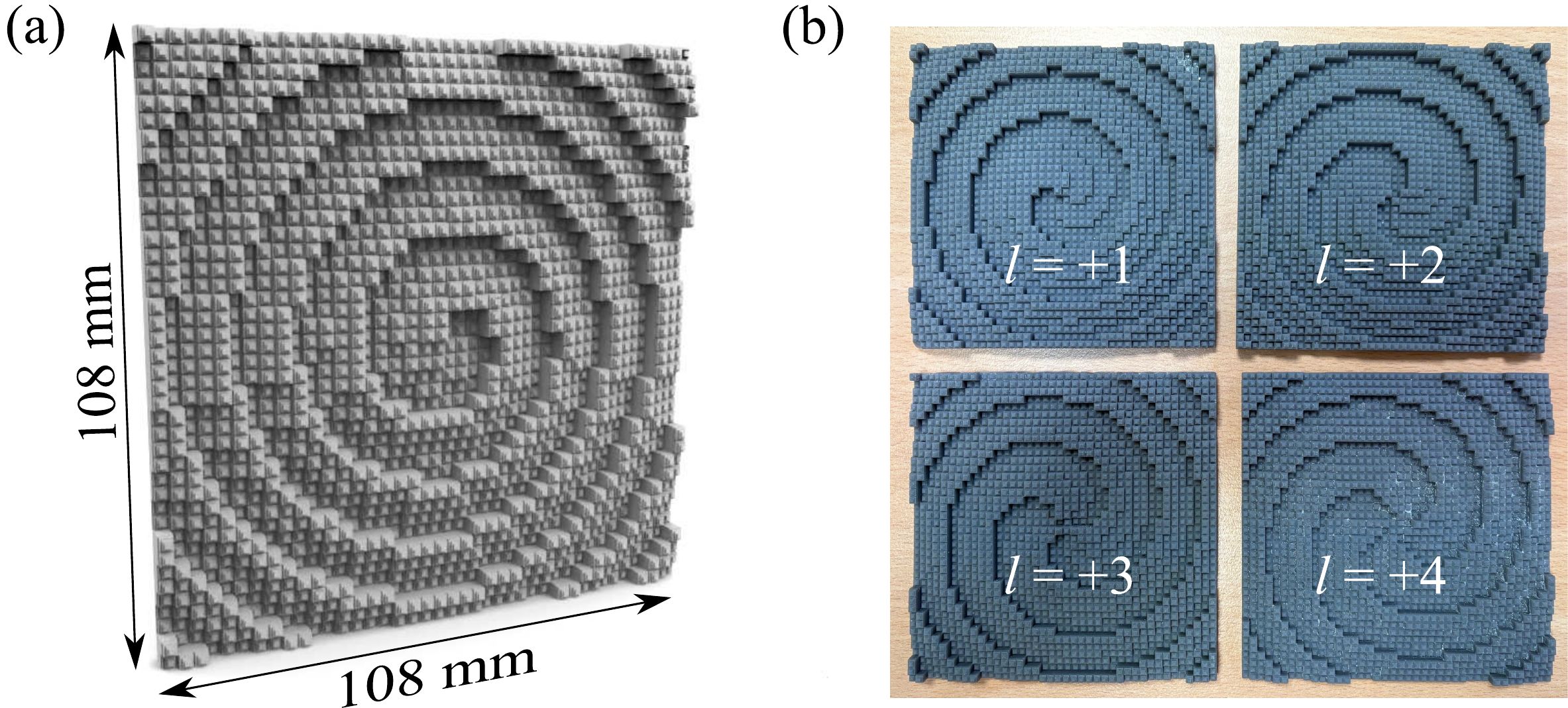}}
    \caption{\footnotesize{(a) Design of $l$=+1 dielectric} TA. (b) Several 3D-printed OAM TAs.}
    \label{Fig:img_OAM}
 \end{figure}

The TA prototypes consist of a $40 \times 40$ array of fully-dielectric unit cells with a periodicity of $p = 2.7\,$ and operating in the Ka-band ($27.5-28.5\,$GHz). They are synthesized by using a 2-dimensional distribution of pyramid unit-cells across the $xy$-plane, forming the dielectric lens architecture, and the $z$-axis accounts for the propagation direction. The TAs were fabricated using stereolithography (SLA) via the FormLabs Form 3 3D printer  using the Grey V4 resin ($\varepsilon_{r} = 2.6$, $\tan \delta = 0.02\,\textrm{at}\,30\,\textrm{GHz}$). An example of TA is depicted in Fig.~\ref{Fig:img_OAM}(a), illustrating the design of a TA with OAM order $l = 1$.

\begin{figure}[]
    \centering
    {\includegraphics[width=0.92\columnwidth]{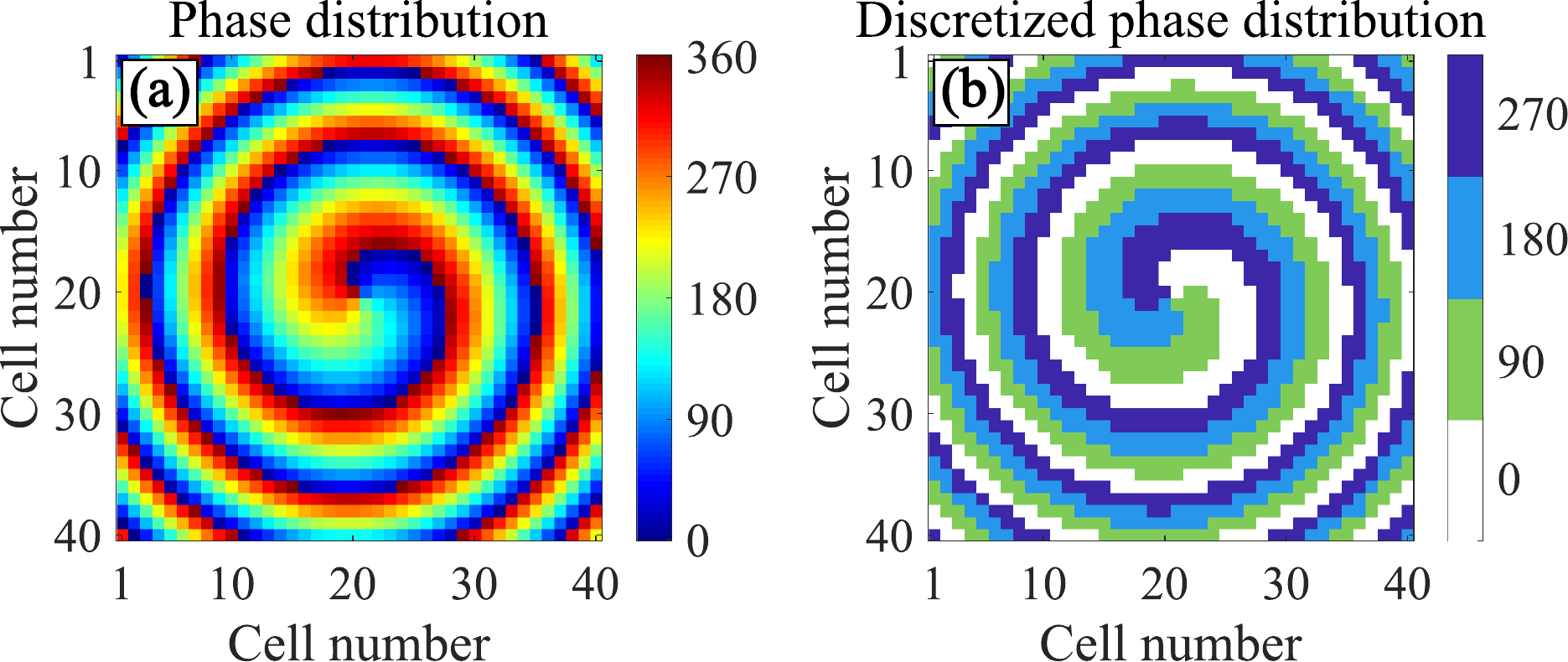}}
    \caption{\footnotesize{(a): Phase-distribution obtained by \eqref{eq:OAM} to generate an OAM order with $l=+1$}. (b): Corresponding discretized TA.  }
    \label{Fig:Discret}
 \end{figure}

A spherical wave is assumed to feed the TAs. The phase $\phi_{mn}$ of the transmitted wave at the output of each individual cell in the position $\mathbf{r}_{mn}$ is determined by the following relationship:
\begin{equation}\label{eq:OAM}
\phi_{mn} = l\varphi_{mn} +k_{0}·\|\mathbf{\mathbf{r}}_{mn} - \mathbf{r}_{\text{f}}\|
\end{equation}
\noindent with $l$ being the order of the desired OAM mode, $\mathbf{r}_{\text{f}}$ the vector from the spherical-wave focus to the center of the TA, and $\varphi_{mn}$ the azimuthal angle of the $(m,n)$-th element. Though \eqref{eq:OAM} is conceived to generate an almost-continuous phase gradient across the TA, in this work we focus on 2-bits configurations for the sake of simplicity in the design. The 2-bits gradient (or 2-bits phase map) discretizes the phase distribution in terms of 4 individual phase states. This discretization is realized according to the following criteria: $\phi_{mn}$ between $0^\circ$ and $90^\circ$ becomes $\phi_{mn} = 0^\circ$; $\phi_{mn}$ between $90^\circ$ and $180^\circ$ becomes $\phi_{mn} = 90^\circ$; $\phi_{mn}$ between $180^\circ$ and $270^\circ$ becomes $\phi_{mn} = 180^\circ$; and $\phi_{mn}$ between $270^\circ$ and $360^\circ$ becomes $\phi_{mn} = 270^\circ$. 
Fig.~\ref{Fig:Discret}(a) shows the phase map in the almost-continuous case for the OAM $l = 1$, while Fig.~\ref{Fig:Discret}(b) denotes the corresponding discretized counterpart. 

\begin{figure}[t]
    \centering
    {\includegraphics[width=1\columnwidth]{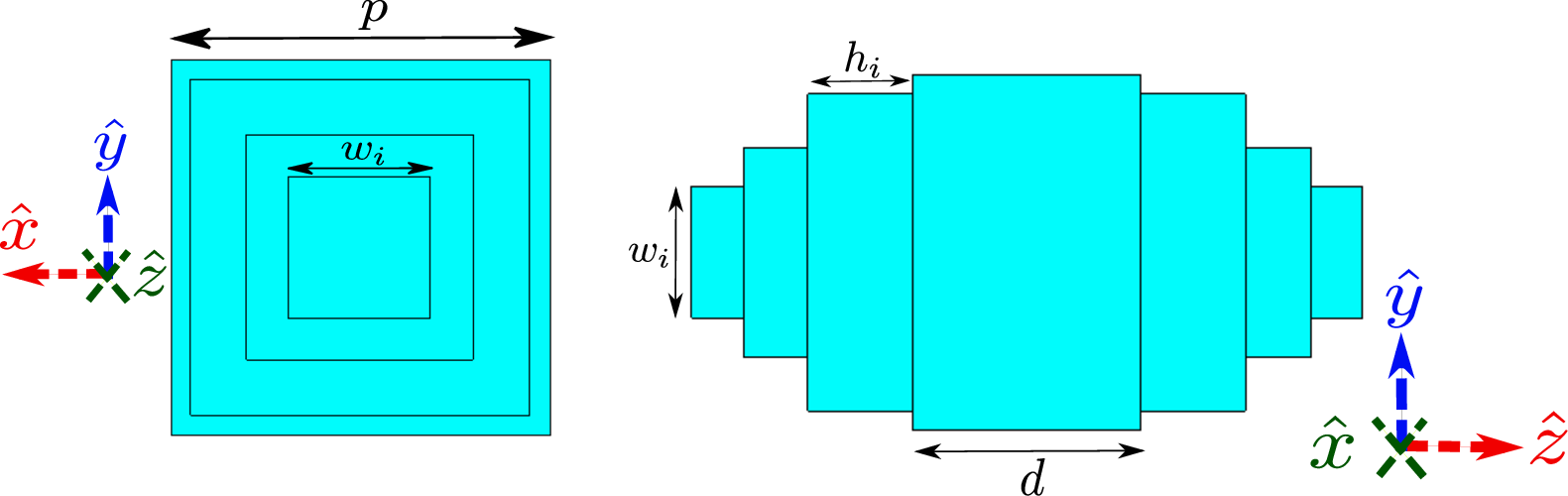}}
    \caption{\footnotesize{A pyramid-shaped unit cell with seven sections;
    $h_1=h_7=0.4\,$mm, $h_2=h_6=0.5\,$mm, $h_3=h_5=0.8\,$mm, $w_1=w_7=1\,$mm, $w_2=w_6=1.6\,$mm, $w_3=w_5=2.4\,$mm, $p=2.7\,$mm, $d$ is the parameter to modify.}}
    \label{fig:unit_cell}
 \end{figure}

The architecture of the corresponding unit cell is depicted in Fig. \ref{fig:unit_cell}. The apparent pyramid-shaped unit cell, studied in \cite{Balma2024-OAM}, consists of the concatenation of square-shaped sections with dimensions $w_{i} \times w_{i} \times h_{i}$ where $i$ refers to the section number of the unit cell, $i = 1 \rightarrow 7$.  
The process for modeling and optimizing this cell is explained in depth in \cite{Balma2024-OAM}.  
The tuning of the phase is governed by changing the width of the central block, $d$. Four cell configurations, whose output phase differ in $90^\circ$, have been designed. The structure parameters are summarized in the caption of Fig.~\ref{fig:unit_cell}. Consequently, TAs going the full range from $l=0$ to $l=+10$ were designed and fabricated. Some examples of the manufactured TAs can be seen in Fig. \ref{Fig:img_OAM}(b), together with their corresponding order.




 \begin{figure}[t]
    \centering
    {\includegraphics[width=0.88\columnwidth]{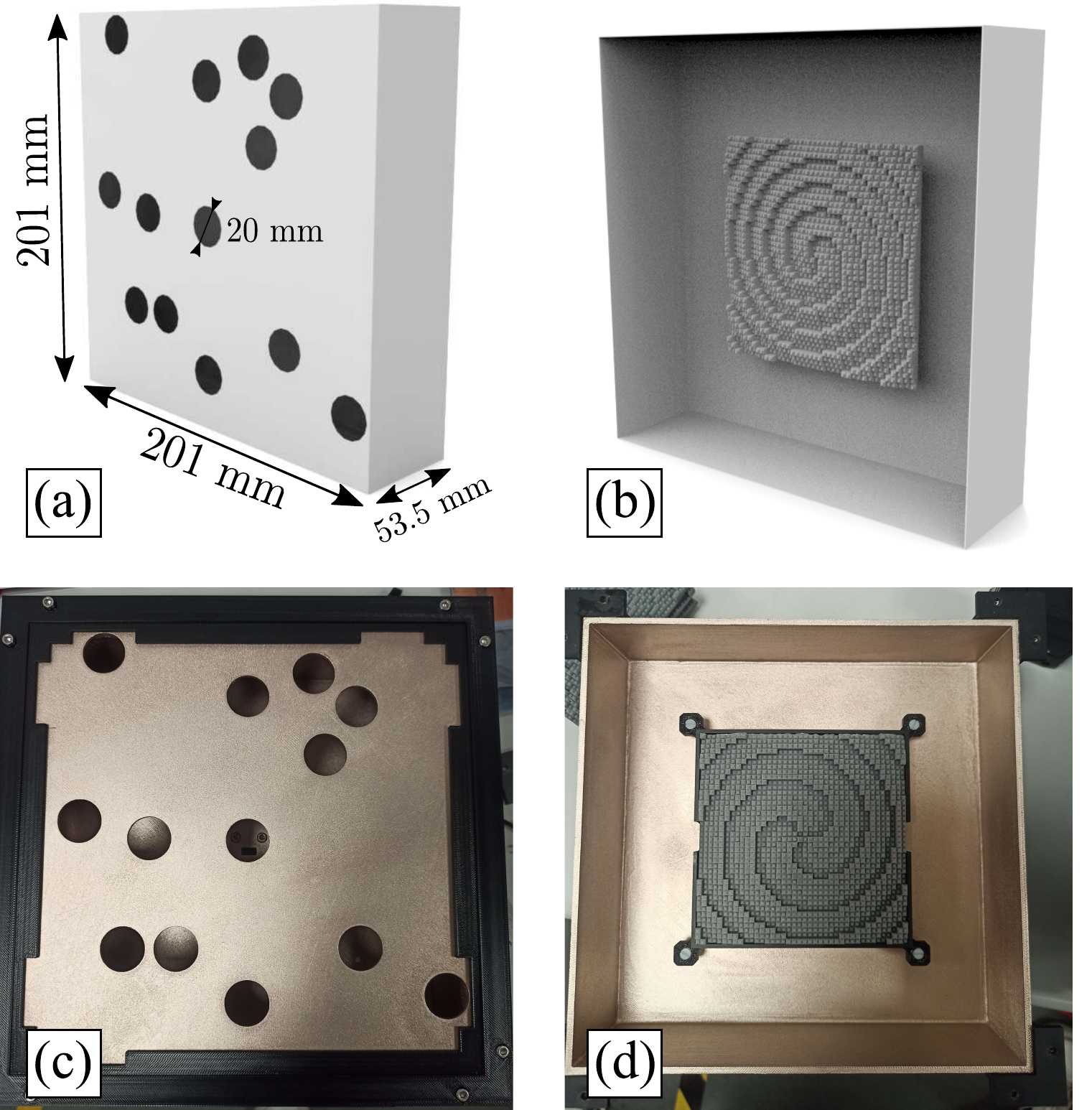}}
    \caption{\footnotesize{(a) Complete metal cavity with its dimensions. (b) Open cavity with the OAM TA ($l$=+3)} inside. (c) Fabricated closed cavity with the random hole pattern lid. (d) Fabricated open cavity with an OAM TA in the locking system.}
    \label{Fig:cavs}
 \end{figure}

\begin{figure}[t]
    \centering
    {\includegraphics[width=0.92\columnwidth]{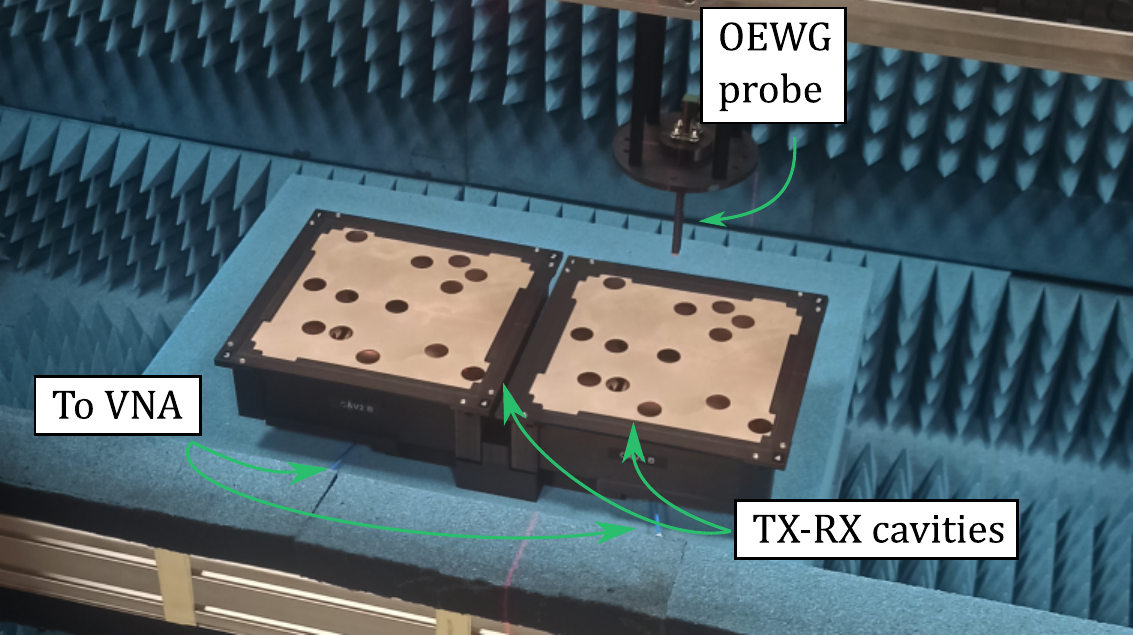}}
    \caption{\footnotesize{TX and RX cavities on the planar antenna measurement range.}}
    \label{fig:setup_NF_scan}
 \end{figure}

 \begin{figure}[t]
    \centering
    {\includegraphics[width=0.93\columnwidth]{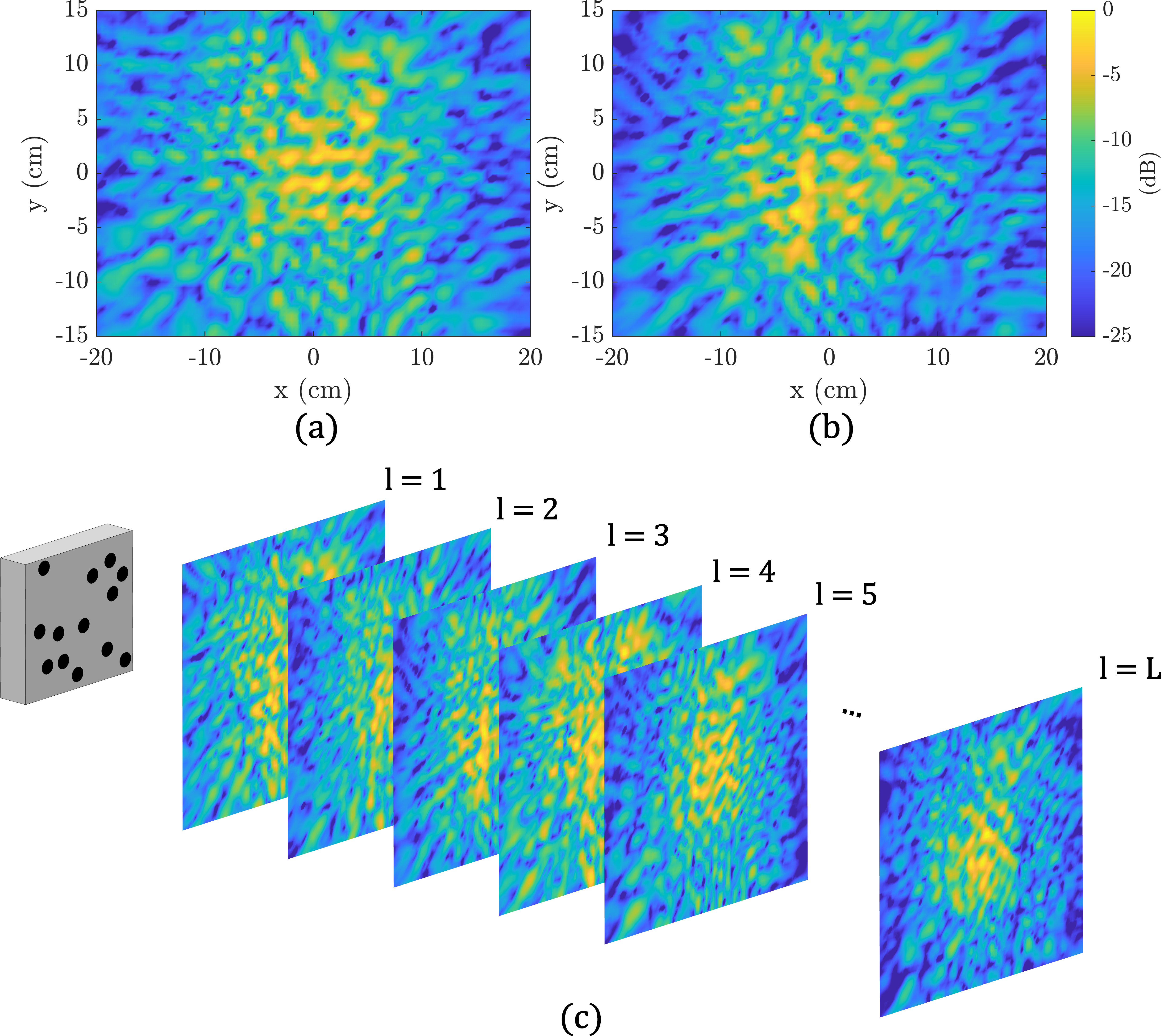}}
    \caption{\footnotesize{Field radiated by one of the cavities measured at $z=15\,\textrm{cm}$ from its aperture when an OAM wave of order (a) $l=1$ and (b) $l=4$} is considered. (c) Diagram of a cavity radiating different patterns considering OAM waves of multiple orders at $f = 28\,$GHz.}
    \label{fig:fields_scheme_OAM}
 \end{figure}
 
\subsection{Tunable Cavity Design}

The cavity developed in this work is depicted in Fig.~\ref{Fig:cavs}. It consists of a square cavity with a leaky front layer that has a series of 13 randomly located holes of $20\,\textrm{mm}$ diameter as radiating elements. The dimensions of the cavity are $201\,\textrm{mm}\times201\,\textrm{mm}\times53.5\,\textrm{mm}$ and is fed by a WR-34 standard waveguide (placed in the center of the back of the cavity), covering the frequency range $22-33\,\textrm{GHz}$. The structure of the cavity is 3D printed using PLA filament and metallized using silver-copper spray paint (with a volume resistivity of $0.00047 \;\Omega\text{cm}$ at $1.7$ mil coat thickness). The OAM TAs presented in Section \ref{sec:TAsdesign} are inserted in the cavity over a non-metallized 3D printed support structure at a distance $\mathbf{r}_{\text{f}} =12.5\,\textrm{mm}$ from the waveguide feeding the cavity, as can be seen in Fig.~\ref{Fig:cavs}(c)-(d), which shows one the fabricated prototypes. In particular, azimuthal-mode orders ranging from $l = +1$ to $l = +10$ in steps of one were considered.




\subsection{Imaging Setup} 

The proof-of-concept CI system used in this work comprises two of the aforementioned cavities arranged side by side as depicted in Fig. \ref{fig:setup_NF_scan}, one acting as TX and other as RX. The distance between the centers of the cavities is $236\,\textrm{mm}$, and the considered frequency range covers from $27.5\,\textrm{GHz}$ to $28.5\,\textrm{GHz}$. If the TAs are not included in the cavities, the arrangement formed by the TX and RX constitutes a frequency diverse CI system. However, this work proposes the use of TAs generating OAM waves inside each cavity to increase the diversity of the measurement modes of the system. In particular, the original number of measurement modes, $N_\text{f}$, is increased by the set of OAM waves generated inside the TX and RX cavities comprising the CI system, noted as $N_{\text{OAM}}$.

The fields radiated by the cavities when each of the considered OAM waves is generated was characterized in the planar antenna measurement range shown in Fig. \ref{fig:setup_NF_scan} using an open-ended waveguide (OEWG) probe. The measured fields radiated by one of the prototypes at $f=28\,\textrm{GHz}$ when two different OAM waves inside the cavity are considered is depicted in Figs.\ref{fig:fields_scheme_OAM}(a)-(b), where the diverse nature of the radiation patterns can be observed. Fig. \ref{fig:fields_scheme_OAM}(c) shows a scheme of the overall operation of the developed prototypes, which reconfigure their radiation pattern employing TAs generating OAM waves of different orders at the same frequency $f = 28\,$GHz.

\section{Results and Discussion}
\label{sec:results}
This section analyzes the impact of considering multiple OAM waves in the performance of a CI system in terms of increased diversity of the measurement modes and improved image quality. For this purpose, the measured fields radiated by the cavities with the TAs generating azimuthal-mode orders ranging from $l = +1$ to $l = +10$ were used to synthesize multiple imaging targets under different conditions.

\begin{figure}[t]
    \centering
    {\includegraphics[width=1\columnwidth]{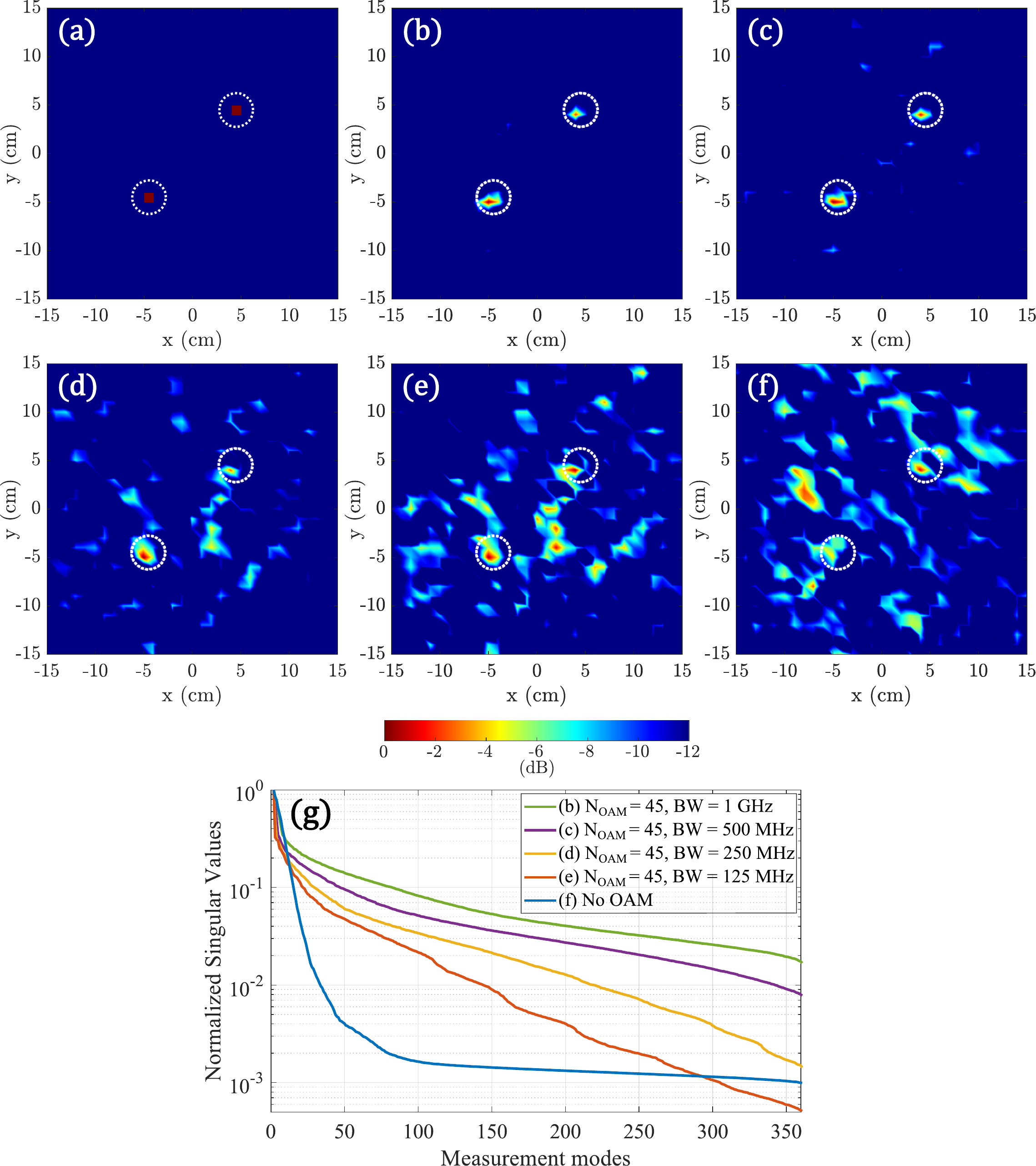}}
    \caption{\footnotesize{(a) Test target and the reconstructed images obtained by the CI system when $N_{\text{OAM}}=45$ and (b) $\textrm{BW}=1\,\textrm{GHz}$, (c) $\textrm{BW}=0.5\,\textrm{GHz}$, (d) $\textrm{BW}=0.25\,\textrm{GHz}$, and (e) $\textrm{BW}=0.125\,\textrm{GHz}$. (f) The same results when only frequency diversity is considered and $\textrm{BW}=1\,\textrm{GHz}$. In all cases, $M = 360$. (g) SVD of the sensing matrix of each configuration.}}
    \label{fig:results_BW}
 \end{figure}

First, a target located at $z=25\,\textrm{cm}$ from the aperture of the cavities comprising two metallic squares of $1\,\textrm{cm}$ side placed at i) $(x_1,y_1)=(-4,-4)\,\textrm{cm}$ and ii) $(x_1,y_1)=(4,4)\,\textrm{cm}$ from the center of the TX/RX arrangement, respectively, was considered [see Fig. \ref{fig:results_BW}(a)]. The image of this target was computed for $f\in[27.5,28.5]\,\textrm{GHz}$ when no OAM waves are considered (i.e., only leveraging the frequency diversity of the cavities, that can be considered as an OAM-wave with order $l = 0$) and for $N_{\text{OAM}}=45$. For a fair comparison, the same number of measurement modes, $M$, is considered in both cases. Thus, the number of frequency points considered within the bandwidth, $N_{\text{f}}$, is adjusted so that $M=N_{\text{f}}\times N_{\text{OAM}}$ remains constant. In particular, $M$ was selected to be $360$, which leads to a separation between the frequency points across the $1\,\textrm{GHz}$ BW of, $\Delta_{f(N_{\text{OAM}}=45)}= 125\,\textrm{MHz}$ and $\Delta_{f\text{(NO-OAM)}}\approx 3\,\textrm{MHz}$. The results are displayed in Figs. \ref{fig:results_BW}(b) and \ref{fig:results_BW}(f) for $N_{\text{OAM}}=45$ and when no OAM modes are considered, respectively. In these images the position of the two squares forming the target is highlighted with a white dotted line. As can be observed, when $N_{\text{OAM}}=45$ (Fig. \ref{fig:results_BW}(b)), the targets are reconstructed without barely any clutter. However, when no OAM waves inside the cavities are considered, Fig. \ref{fig:results_BW}(f), the square of the bottom-left part of the target cannot be imaged and there are multiple areas contaminated with clutter of a magnitude comparable to that of the upper-right target. This means that, considering this bandwidth, the diversity of the radiation patterns generated by the cavities is not sufficient to acquire enough information from the scene. This lack of diversity can be overcome by generating multiple OAM waves inside the cavities [see Fig. \ref{fig:results_BW}(b)], leading to accurate reconstructions of the target. The additional degrees of freedom introduced by the OAM modes yield a significant improvement of the imaging capabilities of the system. Furthermore, the use of multiple OAM modes enable the system BW to be sampled at a greater $\Delta_f$ while ensuring the same total number of measurement modes, $M$. As a result of the larger separation between frequency measurements, the correlation between them is reduced.

To analyze the impact of reducing the operational bandwidth of the system, the image of the same target displayed in Fig. \ref{fig:results_BW}(a) is computed for $N_{\text{OAM}}=45$ and $\text{BW}=0.5\,\textrm{GHz}$, $\text{BW}=0.25\,\textrm{GHz}$, and $\text{BW}=0.125\,\textrm{GHz}$. If the BW is reduced by half, i.e., $\text{BW}=0.5\,\textrm{GHz}$ [Fig. \ref{fig:results_BW}(c)], keeping $M$=360, $\Delta_f$ is $\approx63\,\textrm{MHz}$. Thus, the separation between the frequency measurements across the swept BW is reduced by half. In this case, there is a minor increase in the clutter present in the image, but its quality remains high. When only a quarter of the BW is used [Fig. \ref{fig:results_BW}(d)], $\Delta_f$ is reduced to $\approx31\,\textrm{MHz}$ and, although there is a significant increase of the clutter level, the reflectivity of the targets remain slightly above the clutter level. Finally, if the BW is further reduced to only $125\,\textrm{MHz}$ ($\Delta_f\approx16\,\textrm{MHz}$ to ensure $M=360$), Fig. \ref{fig:results_BW}(e), the target is still reconstructed, but the image is heavily contaminated with clutter which, in some areas, presents the same magnitude as that of the areas corresponding to the target.

The diversity of the measurement modes for each configuration can be assessed by computing the singular value decomposition (SVD) of the sensing matrix of the resulting CI system. The flatter the SVD spectrum, the more diverse are the measurement modes of the CI system, indicating that more information from the inspected scene can be recovered. The SVDs corresponding to each analyzed configuration are depicted in Fig. \ref{fig:results_BW}(g). As can be observed, when a pure frequency diverse system is considered (dark blue line), the magnitude of the singular values drops fast, meaning that only a few measurement modes contribute effectively to the reconstruction of the imaging scene. In contrast, when $N_{\text{OAM}}=45$, the SVD is flatter. Furthermore, as the BW increases, the flatness of the SVD consistently increases, in agreement with the evolution of the quality of the images from the target displayed in Fig. \ref{fig:results_BW}.

\begin{table}[t]  
\scriptsize
\caption{TCR of the images of Fig. \ref{fig:results_BW}.}
\label{table:BW_TCR}
\setlength{\tabcolsep}{4pt}
\renewcommand{\arraystretch}{1.35}
\centering %
\begin{tabular}[t]{>{\centering}p{0.15\linewidth}>{\centering}p{0.1\linewidth}>{\centering}p{0.10\linewidth}>{\centering}p{0.1\linewidth}>{\centering}p{0.1\linewidth}>{\centering}p{0.03\linewidth}>{\centering\arraybackslash}p{0.15\linewidth}}
\toprule
    & \multicolumn{4}{c}{$N_{\text{OAM}}=45$} &    & No OAM \\
    \cmidrule{2-5}     
BW (GHz)			&	1	&	     0.5		&	0.25	&	0.125  &	&	1	\\
\midrule
    TCR (dB) 			&		$20.6$		&		$18.3$		&		$14.8$		&		$13.7$	&							&	$10.6$		\\
\bottomrule
\end{tabular}
\end{table}

To compare quantitatively the quality of these images, their target-to-clutter ratio (TCR) \cite{freehand_MIMO} was computed. The results are summarized in Table \ref{table:BW_TCR}, where it can be observed that the TCR is higher when $N_{\text{OAM}}=45$ for all the analyzed BWs, than when no OAM waves are considered. Furthermore, for a constant $N_{\text{OAM}}$, the quality of the reconstructed images decreases as the BW is reduced, in agreement with the previous discussion from a qualitative point of view. Looking at these results, it can be concluded that without considering multiple OAM waves, not even with $1\,\textrm{GHz}$ of BW the system is capable of reconstructing the target. In contrast, when $N_{\text{OAM}}=45$, it is possible to drastically reduce the operational BW (to only $250\,\textrm{MHz}$) and still obtain an accurate image of the scene under inspection. It should be noted that a tradeoff between the number of considered OAM waves, $N_{\text{OAM}}$, and the BW required to enable the accurate reconstruction of the scene must be established. Thus, if the available BW is significantly reduced, a higher $N_{\text{OAM}}$ can be considered.

\begin{figure}[t]
    \centering
    {\includegraphics[width=1\columnwidth]{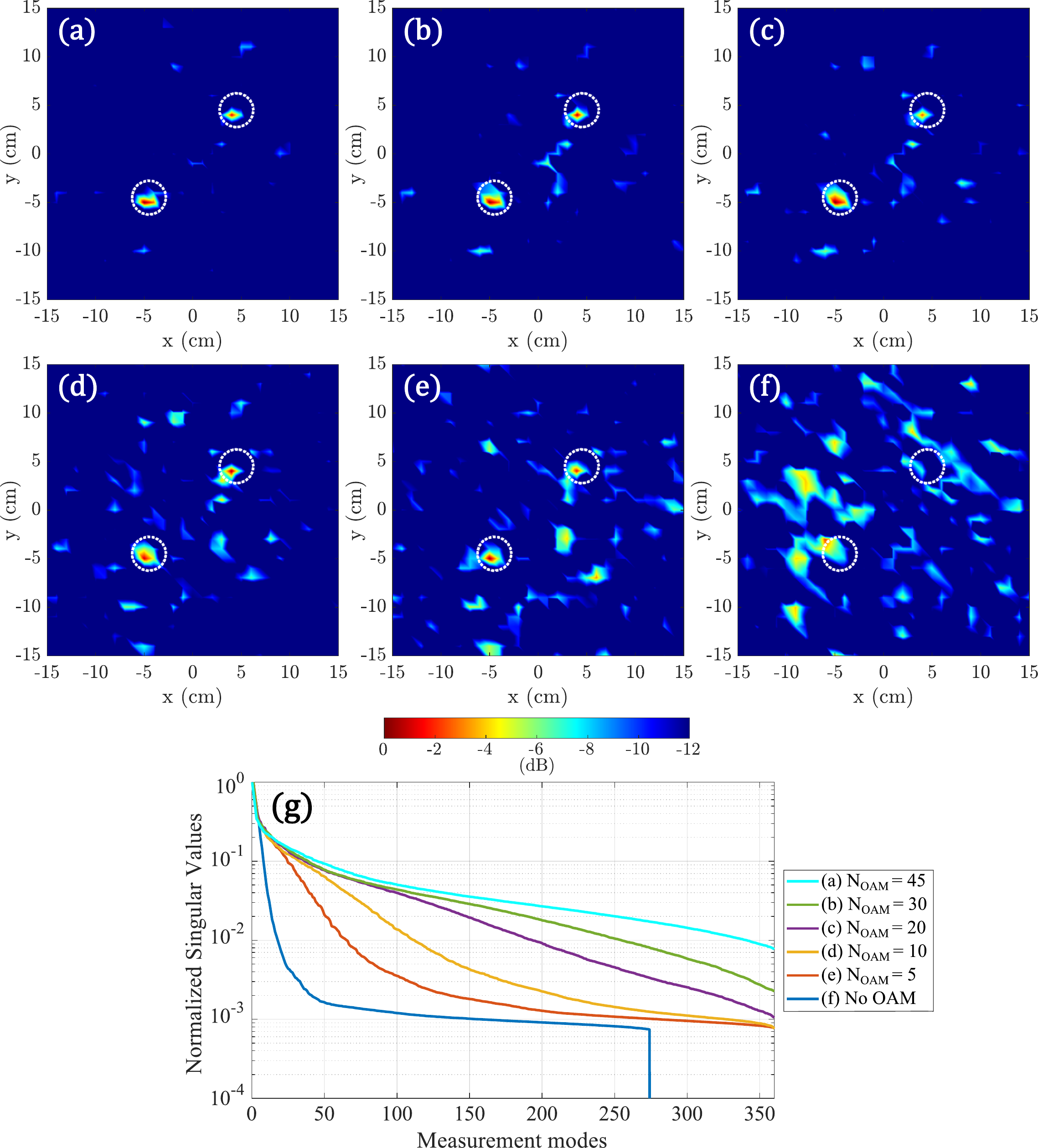}}
    \caption{\footnotesize{Reconstructed images obtained by the CI system when (a) $N_{\text{OAM}}=45$, (b) $N_{\text{OAM}}=30$, (c) $N_{\text{OAM}}=20$, (d) $N_{\text{OAM}}=10$, (e) $N_{\text{OAM}}=5$, and (f) only frequency diversity is considered. In all cases, $M = 360$. (g) SVD of the sensing matrix of each configuration.}}
    \label{fig:results_NOAM}
 \end{figure}

\begin{table}[t]  
\scriptsize
\caption{TCR of the images of Fig. \ref{fig:results_NOAM}.}
\label{table:NOAM_TCR}
\setlength{\tabcolsep}{4pt}
\renewcommand{\arraystretch}{1.35}
\centering %
\begin{tabular}[t]{>{\centering}p{0.15\linewidth}>{\centering}p{0.095\linewidth}>{\centering}p{0.095\linewidth}>{\centering}p{0.095\linewidth}>{\centering}p{0.095\linewidth}>{\centering}p{0.095\linewidth}>{\centering\arraybackslash}p{0.11\linewidth}}
\toprule
$N_{\text{OAM}}$			&	45	&	     30		&	20		&	10  &	5		&	No OAM\\
\midrule
	TCR (dB) 			&		$18.3$		&		$17.4$		&		$16.3$		&		$15.3$	&			$14.0$					&	$6.4$		\\
\bottomrule
\end{tabular}
\end{table}

Next, the impact of increasing the number of considered OAM waves is assessed. In this case, the same configuration used to obtain Fig. \ref{fig:results_BW}(c) was employed, i.e., $M=360$ and $\text{BW}=0.5\,\textrm{GHz}$. In particular, the image of the same target is computed for $N_{\text{OAM}}=45$, $N_{\text{OAM}}=30$, $N_{\text{OAM}}=20$, $N_{\text{OAM}}=10$, $N_{\text{OAM}}=5$, and without OAM waves. As in the previous analysis, to keep the same number of measurement modes, $M$, the value of $\Delta_{f}$ changes for each configuration. Specifically, the separation between the frequency points across the BW is, $\Delta_{f(N_{\text{OAM}}=45)}\approx 63\,\textrm{MHz}$, $\Delta_{f(N_{\text{OAM}}=30)}\approx 42\,\textrm{MHz}$, $\Delta_{f(N_{\text{OAM}}=20)}\approx 28\,\textrm{MHz}$, $\Delta_{f(N_{\text{OAM}}=10)}\approx 14\,\textrm{MHz}$, $\Delta_{f(N_{\text{OAM}}=5)}\approx7 \,\textrm{MHz}$, $\Delta_{f(\text{NO-OAM})}\approx 1\,\textrm{MHz}$, respectively.

As previously discussed, when $N_{\text{OAM}}=45$ [Fig. \ref{fig:results_NOAM}(a)], the target is accurately reconstructed and the clutter level is low. When $N_{\text{OAM}}$ is reduced to $30$, $20$ and $10$, Figs. \ref{fig:results_NOAM}(b), \ref{fig:results_NOAM}(c) and \ref{fig:results_NOAM}(d), respectively, the target can still be imaged, although the clutter level progressively increases. If $N_{\text{OAM}}$ is further reduced to $5$, Fig. \ref{fig:results_NOAM}(e), clutter of a similar magnitude to that of the target appears in some parts of the image. Finally, if no OAM waves inside the cavities are considered, as shown in Fig. \ref{fig:results_NOAM}(f), the target cannot be reconstructed. The SVD spectrum of these configurations is illustrated in Fig. \ref{fig:results_NOAM}(g). In agreement with the previous discussion, the fastest decay of the SVD spectrum takes place when no OAM waves are considered. Furthermore, the SVD spectrum becomes flatter as $N_{\text{OAM}}$ increases. This analysis confirms that, considering the same number of measurement modes employing a given operational BW, the imaging capabilities of the system benefit from the use of OAM modes as an additional degree of freedom to increase its diversity.


To compare quantitatively the quality of these images, their TCR was computed. The obtained values, which are in accordance with the previous discussion, are summarized in Table \ref{table:NOAM_TCR}. The highest TCR is associated with $N_{\text{OAM}}=45$ and the flatter SVD spectrum. In addition, the TCR decreases with $N_{\text{OAM}}$ and drops significantly when only frequency diversity (no OAM) is considered to generate the measurement modes of the CI system.

\begin{figure}[t]
    \centering
    {\includegraphics[width=1\columnwidth]{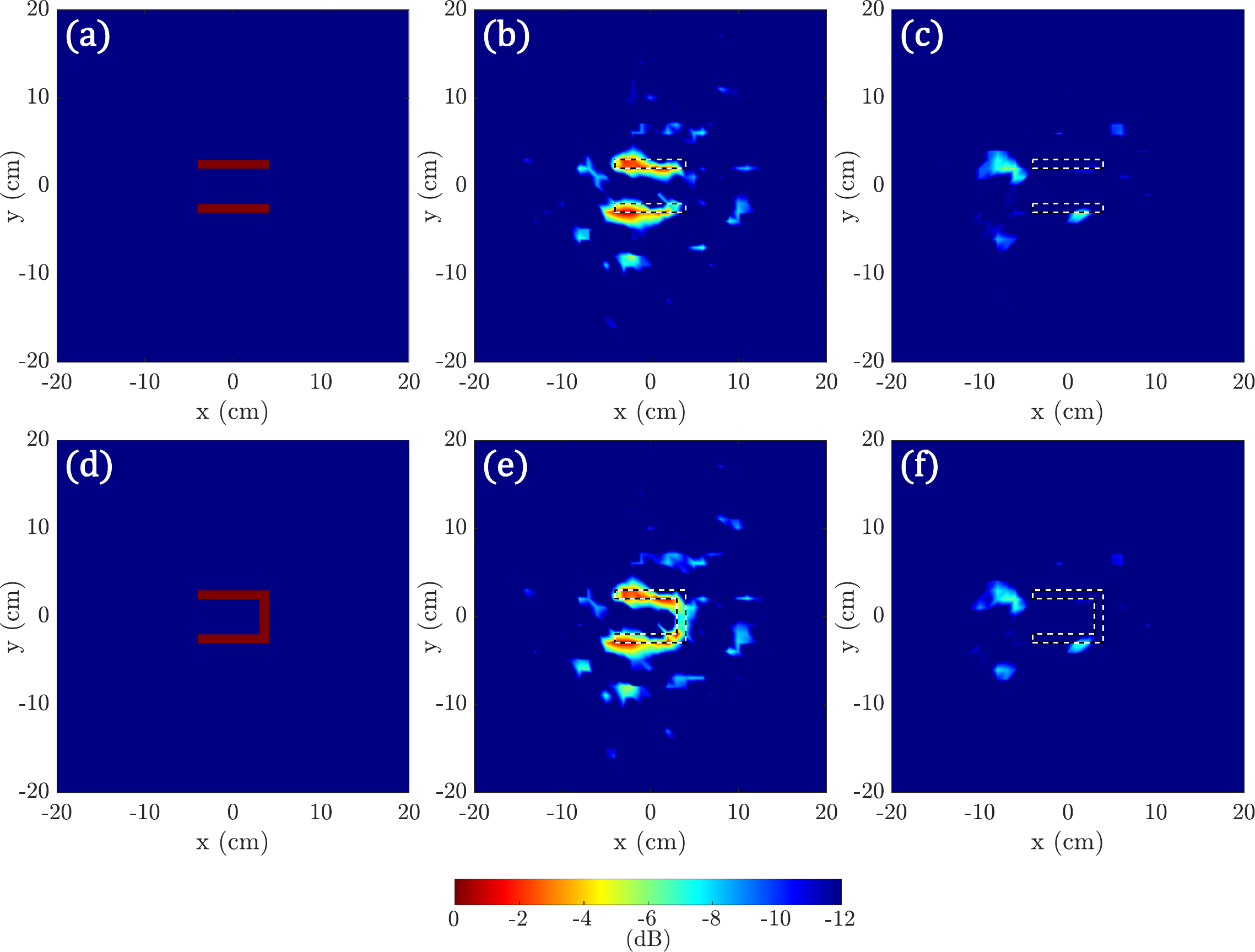}}
    \caption{\footnotesize{Target comprising two parallel strips along the $x$ axis: (a) ground truth, and reconstructed image for (b) $N_{\text{OAM}}=45$ and (c) when no OAM waves are considered. ``U'' shaped target: (d) ground truth, and reconstructed image for (e) $N_{\text{OAM}}=45$ and (f) when no OAM waves are considered.}}
    \label{fig:results_complex}
 \end{figure}

Following the detailed CI analyses presented for the square patch targets presented in Fig. \ref{fig:results_BW} and Fig. \ref{fig:results_NOAM}, a series of additional CI studies of more complicated, distributed set of targets were carried out. To this end, two additional targets were imaged to further assess the capabilities of the proposed system. In particular, a target comprising two metallic parallel $8\,\textrm{cm}$ long, $1\,\textrm{cm}$ wide strips separated $4\,\textrm{cm}$ (Fig. \ref{fig:results_complex}(a)), and the ``U'' shaped target depicted in Fig. \ref{fig:results_complex}(d). Both targets were reconstructed for $BW=1\,\textrm{GHz}$ when $N_{\text{OAM}}=45$ (Figs. \ref{fig:results_complex}(b) and \ref{fig:results_complex}(e)), and when no OAM waves are considered (Figs. \ref{fig:results_complex}(c) and \ref{fig:results_complex}(f)). The dashed black and white line indicates the contour of the targets in Figs. \ref{fig:results_complex}(b)-(c) and Figs. \ref{fig:results_complex}(e)-(f). As can be observed, when $N_{\text{OAM}}=45$ (Figs.\ref{fig:results_complex}(b) and (e)) the targets can be successfully reconstructed and the clutter level of the images is low. In contrast, when only the frequency diversity of the cavity is considered (i.e., no OAM waves are used), the imaging system is not capable of successfully retrieving the image of the targets (Figs.\ref{fig:results_complex}(c) and (f)).


 \vspace{-1pt}

\section{Conclusion}
\label{sec:conclusion}
This work investigated the image reconstruction performance of a CI system composed of two perforated cavities operating as TX and RX. The novelty of the approach lies in the introduction of TAs configured for various OAM orders. This implementation introduced a flexible framework that enables multiple combinations of TX and RX configurations across different OAM orders, resulting in a substantial increase in modal diversity without requiring excessive BW. Imaging results demonstrated that the proposed system performs effectively when considering several OAM modes (even for a reduced frequency BW), whereas for the empty cavities a higher BW would be needed to achieve similar diversity. In particular, when considering $N_{\text{OAM}}=45$ modes, the targets were accurately reconstructed even with a bandwidth as narrow as $250\,\text{MHz}$, whereas for the empty cavities the targets cannot be distinguished even with a $1\,\textrm{GHz}$ BW. Various target geometries have been imaged, and the reconstruction results confirm the system’s capabilities. These preliminary results demonstrate the efficiency of combining OAM waves and perforated frequency-diverse cavities, and pave the way for future conceptions based on faster reconfigurable TAs for next generation CI systems.




\bibliographystyle{IEEEtran}
\bibliography{bibliography}

\end{document}